\documentstyle[aps,prl,multicol,epsfig,psfig]{revtex}
\begin{document}

\title{Kosterlitz-Thouless vs Ginzburg-Landau description of
2D superconducting fluctuations} 
\author{L. Benfatto, A. Perali, C. Castellani, and M. Grilli} 

\address{Dipartimento di Fisica, Universit\`{a} di Roma ``La
Sapienza'' and \\ 
Istituto Nazionale Fisica della Materia, Unit\`a di Roma 1,\\
P.le A.  Moro, 2 - 00185 - Roma, Italy}

\date{\today} 

\maketitle 
\medskip

\begin{abstract}
We evaluate the
charge and spin susceptibilities of the 2D attractive Hubbard model and
we compare our results with Montecarlo simulations on the same model.
We discuss the possibility to include topological 
Kosterlitz-Thouless superconducting fluctuations
in a standard perturbative approach substituting in the fluctuation 
propagator the Ginzburg-Landau correlation length with the 
Kosterlitz-Thouless correlation length.
\end{abstract}

\vskip 2pc 

\begin{multicols}{2}

The discovery of spin and charge pseudogaps in the normal state of
underdoped superconducting cuprates \cite{timusk}
has triggered a renewed interest on the physics of preformed 
Cooper pairs. The actual source of the pseudogaps (pairing, and/or spin-,
and/or charge fluctuations) and the leading mechanisms responsible for 
the reduction of the superfluid density at low temperature
(classical phase fluctuations, collective modes, 
quasiparticle excitations) are still debated. 
However, many indications support the idea that pairing occurs
below some crossover temperature $T^*$, while the phase coherence
is established at a sizable lower temperature. 
The low density of carriers resulting  in a low superfluid density and
the short coherence length $\xi_0\sim 10\div 20\AA$,
support the relevance of the superconducting phase fluctuations 
in the thermodynamic and dynamic properties of these materials.
Moreover, although no discontinuity of the superfluid density 
at $T_c$ is observed, the strong anisotropy of the cuprates
suggests that some features of a Kosterlitz-Thouless (KT)
transition could be present in these systems \cite{corson}. 
Therefore it is worth investigating the effects of the topological
vortex-antivortex phase fluctuations on the various properties of a 2D 
superconductor. In particular, an important issue concerns the inclusion of 
these effects in evaluating thermodynamic quantities like the spin 
susceptibility or the charge susceptibility.
In this context, the aim of the present work is to look for
possible connections between the perturbative scheme leading to the standard 
time-dependent Ginzburg-Landau (TDGL) results and the KT physics. 

Halperin and Nelson \cite{Halperin} have shown that, in the
KT regime, the contributions of superconducting fluctuations to the
conductivity above $T_{KT}$
have the same functional form, in terms of the correlation length $\xi$,  
as the Aslamazov-Larkin contributions of the standard TDGL theory,
$\sigma _{KT}(\xi) \simeq \sigma _{GL}(\xi)\sim \xi ^2 $.
The same holds for the fluctuation contribution to the diamagnetism
$\chi ^{d}_{KT}(\xi)\simeq \chi ^{d}_{GL}(\xi)\sim \xi ^2$.
In spite of the same correlation length dependence, conductivity and 
diamagnetism in KT or TDGL theory  have a completely different 
temperature dependence, induced by the different temperature 
dependence of the correlation length in the two theories. 
The KT correlation length diverges exponentially at $T_{KT}$ 
while the GL correlation length 
diverges as a power-law with the classical exponent $\nu =\frac{1}{2}$. 
Therefore the KT conductivity and the diamagnetic susceptibility
diverge exponentially at $T_{KT}$ while the same quantities
in the TDGL theory diverge as a power-law at 
$T_c$ with a critical exponent $\gamma =1$.
In the present work we investigate the possibility that,
in analogy with conductivity and diamagnetism, the correct behavior 
of the spin and charge susceptibilities in the KT regime can be simply 
recovered by inserting the KT correlation length 
in their TDGL expressions. 
We shall find that this prescription does work for the spin susceptibility 
while it does not for the charge susceptibility.

We analyze the two-dimensional negative-$U$ 
Hubbard model \cite{Micnas} which
is the simplest minimal model where the distinct
occurrence of pairing and phase coherence can be investigated. 
Within this model, 
the spin susceptibility $\chi_s$ and the charge compressibility $\chi_c$
are calculated  on a two-dimensional square lattice
by performing a loop expansion with the fermions exchanging
the Cooper-fluctuations propagator in the standard form. 
Before giving the technical details of our treatment, 
we immediately present our results. 

Figure 1 shows the behavior of the spin susceptibility when the correlation
length is assumed either of the GL form (dashed line with crosses) 
or of the KT form (dotted line with stars). Both curves are compared with
the Montecarlo data obtained in Ref.\onlinecite{Randeria,Singer} for the 
negative-$U$ Hubbard model with $U=-4t$ 
($t$ is the nearest-neighbor hopping)
at filling $n=0.5$ electrons per cell. The critical temperature $T_{KT}$
of the KT superconducting transition, as extracted from numerical
calculations, is $T_{KT}=0.05t$ and has been used 
as the input critical temperature for our perturbative calculations. 
In the Montecarlo data, for $T$ less then 
$T^\star \simeq t \gg T_{KT}$, 
$\chi_s$ starts decreasing. This indicates the
existence of strong superconducting fluctuations in the temperature
range between the mean-field transition temperature ($T_{BCS}\simeq 0.6t$) 
and the true KT transition.                     
It is apparent from Fig. 1 that the rapid decrease of the spin 
susceptibility in the Montecarlo results is well fitted by 
inserting in the correlation length the KT temperature dependence as 
given by the expression 
\begin{equation}
\label{correlation}
\xi_{KT}(T)=\xi_c\exp
\left [ b\sqrt{\frac{T(T_{BCS}-T_{KT})}{T_{BCS}(T-T_{KT})}}\right].
\end{equation}
Here $\xi_c$ is an effective size of the core of the vortex that we 
take of the order of the zero temperature correlation length $\xi_0$, 
and $b$ is a positive constant of the order of unity.
This specific form of the KT correlation length has been derived along the 
line of Ref.\cite{Halperin}, although it
differs slightly from the one commonly 
quoted in the literature \cite{KT1,KT2}.
We shall comment on this later.
Notice that the KT mass term (inverse square of the correlation length) 
of the Cooper propagator
remains small and  generates strong fluctuations, 
in a wider range of temperatures than the GL mass with the same
critical temperature in agreement with Montecarlo data. 
The GL correlation length is instead completely inadequate to 
reproduce the Montecarlo data in the all range of temperatures.

\begin{figure}
\narrowtext
\psfig{figure=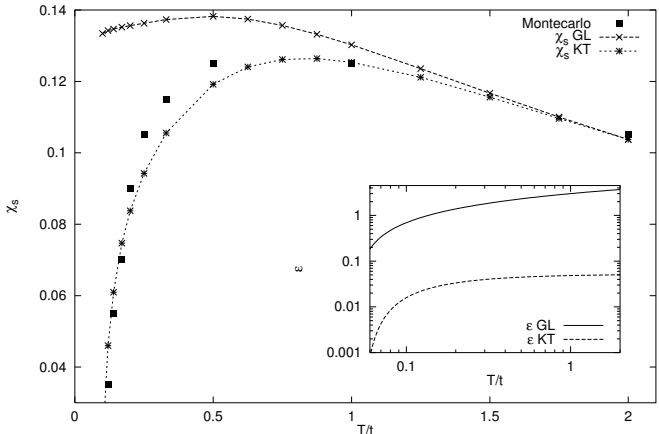,width=6cm,angle=-90}
\caption{Comparison between  Montecarlo spin susceptibility 
(taken from Ref. [6]) and
the spin susceptibility calculated using the Ginzburg-Landau 
($\chi_s^{GL}$) and the Kosterlitz-Thouless ($\chi_s^{KT}$) 
correlation length.}
\label{fitspin}
\end{figure}

The fit in Fig. 1 stops at $T\simeq 0.1t$ because 
there are no numerical data below this value. This also appears to be 
the lower limit for our approach to work. 
Indeed for $T\simeq 0.09t$ the TDGL expression for $\chi_s$ develops a 
non physical behavior $(\chi_s < 0)$, 
indicating that the perturbative scheme no longer applies near $T_{KT}$. 
Whit this caution in mind, the results of Fig. 1 indicate that the simple
loop expansion we adopted is able to reproduce the  spin susceptibility 
in a wide range of temperatures.  They support the idea that
the main effect of the vortex-antivortex phase fluctuations on the spin 
susceptibility is embedded in 
(and satisfactorily accounted for by)
the temperature dependence of the $\xi_{KT}(T)$ correlation length, in 
analogy with the conductivity and diamagnetism.

On the other  hand, as seen in Fig. 2, 
the same type of calculations for the charge
susceptibility fail in describing the nearly constant (but with
sizeable error bars) behavior obtained numerically. In particular, 
we find that the Aslamazov-Larkin (AL) contribution, which does not 
contribute to the spin susceptibility, strongly enhances $\chi_c(T)$
and eventually leads to a divergent $\chi_c$ near $T_{KT}$. 
As a consequence $\chi_c(T)$ strongly deviates from the 
Montecarlo results for $T<T_{BCS}$.
In Fig. 2 we also report the RPA resummation of the bare bubble in the
charge channel that fits the available Montecarlo data, to obtain,
by extrapolation, the $\chi_c(T)$ at higher temperature.

\begin{figure}
\narrowtext
\psfig{figure=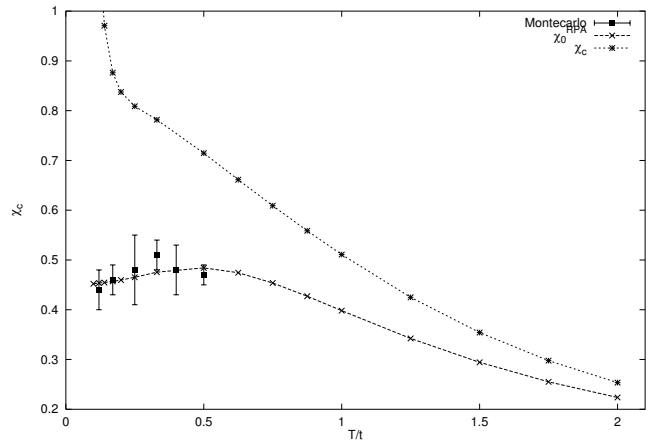,width=6cm,angle=-90}
\caption{Comparison between  Montecarlo charge susceptibility 
(taken from Ref. [8]),
the charge susceptibility calculated using the Kosterlitz-Thouless 
correlation length ($\chi_c$) and the RPA resummation of the bare
bubble ($\chi_{0}^{RPA}$).}
\label{fitcharge}
\end{figure}

With this respect $\chi_c$ appears to behave as the 
specific heat $c_v$, for which the 2D-TDGL expression 
$c_v \sim \xi_{GL}^2$ does not reproduce the correct 
KT result $c_v \sim \xi_{KT}^{-2}$, even when expressed in terms of the 
correlation length. For the specific heat this happens 
despite the free energies in the two theories 
have the same leading behavior 
when written in terms of the respective correlation lengths: 
$F_{GL}\sim \xi_{GL}^{-2}\ln\xi_{GL}$ 
and $F_{KT}\sim \xi_{KT}^{-2}$ \cite{KT1,KT2}.
Indeed, since $c_v$ involves the second derivative of $F$ with 
respect to temperature the different temperature dependences of the 
correlation lengths (and the subleading $\ln\xi_{GL}$ factor) 
lead to completely different results in the two theories. 
Our result for $\chi_c$ has the same origin:
The charge response at $\omega=0, q\rightarrow 0$ 
can be obtained as a chemical
potential derivative of the free energy. Now, since 
the critical temperature depends on the chemical potential
$T_c=T_c (\mu)$, a total derivative with respect to $\mu$ also
involves derivatives with respect to $T_c$, and, in turn, 
derivatives of $\xi$. Therefore the temperature
dependence in $\chi_c$ not only arises from the temperature
dependence of $\xi (T)$, but also depends on $d\xi/dT_c$. In fact one gets 
the same TDGL singular contribution $\sim \xi^2$ for $\chi_c$ and $c_v$.
Our simple perturbative expansion, where the leading
temperature dependence only arises from the mass term $\xi^{-2}$ of the 
Cooper fluctuation propagator in the TDGL 
expression, fails to reproduce the
correct temperature dependence for $\chi_c$ in the same way as it fails 
in evaluating the specific heat.

We now describe the details of our calculations. The model we consider 
is given by 
\begin{equation}
H=-t\sum_{<i,j>\sigma}c^{\dagger}_{i\sigma}c_{j\sigma}
+U\sum_{i}n_{i\uparrow}n_{i\downarrow}-\mu\sum_{i\sigma}n_{i\sigma}
\label{mod}
\end{equation}
where $t$ is the hopping between 
nearest-neighbor sites, $U<0$ the strength
of the attraction and $\mu$ the chemical potential.
The standard ladder resummation of diagrams leads to the Cooper pair
propagator $L(q,\Omega _l)=-U/\left(1+U\chi_0^{pp}(q,\Omega _l)\right)$ 
where $\chi_0^{pp}(q,\Omega _l)$ is the bare particle-particle bubble,
being $q$ the momenta and $\Omega_l$ the Matsubara frequency. 
In the normal state, within the standard GL approach, at 
small $q$ and $\Omega _l$ one has

\begin{equation}
\label{prop}
L^{-1}(q,\Omega_l)=
N_0\left( \epsilon+\eta q^2 + \gamma\mid \Omega _l\mid \right)
\end{equation}
where $N_0$ is the density of states at the Fermi energy,
$\eta=7\zeta(3)/(32\pi^2)(v_F/T_c)^2 \simeq \xi_0^2$, 
and $\gamma=\pi/(8T_c)$. 
The mass term $\epsilon =\ln(T/T_c)=(\xi/\xi_0)^{-2}$ 
of the propagator controls the distance from the 
superconducting transition.  In the standard 
GL approach $\epsilon \sim \xi_{GL}^{-2}$ and near $T_c$ it goes 
to zero as $(T-T_c)/T_c$. 

We study the charge and
spin susceptibilities by evaluating the one loop corrections 
$\Delta \chi _{c}$ (charge channel) and $\Delta \chi _{s}$ (spin channel)
to the bare particle-hole bubble 
$\chi_0^{ph}$, $\chi_{c,s}^{ph}=\chi_0^{ph}+\Delta \chi_{c,s}$.
The charge $(c)$ and spin $(s)$ bubbles $\chi _{c,s}^{ph}$
are then inserted in the RPA resummation to get the charge and spin
susceptibilities (see below).
In the one loop expansion, we include diagrams containing only 
one integration on the bosonic variables $(q,\Omega _l)$
($i.e.$ one bosonic loop) of the fluctuation propagator $L(q,\Omega _l)$, 
obtaining three kinds of diagrams which contribute differently
to the spin and charge
susceptibilities: the selfenergy diagrams, where $L(q,\Omega _l)$
renormalizes the one 
particle bare Green function (DOS contribution); the vertex diagrams, 
where $L(q,\Omega _l)$ renormalizes
the vertex, connecting two bare Green function
(Maki-Thompson (MT) contribution); the Aslamazov-Larkin (AL)
diagrams, containing two fluctuation propagators. 
Moreover it is necessary to add the counterterms (CT) proportional 
to the shift of the chemical potential $\delta \mu$, which is required
to preserve the number of particles. 
We notice that the one loop expansion for the charge
and the spin susceptibilities satisfies the relation, 
derived from spin and charge conservation,
$\chi_{s,c} (q=0, \Omega \neq 0)=0$.
One obtains:
\begin{eqnarray}
\Delta \chi_s&=&4\cdot DOS - 2 \cdot MT + 4 \cdot CT\\
\Delta \chi_c&=&4\cdot DOS + 2 \cdot MT + 4 \cdot AL + 4 \cdot CT.
\end{eqnarray}

The absence of the AL contribution and the (opposite) sign of the MT
diagrams in the spin susceptibility is the consequence of the
vertex spin structure, as shown in Ref. \cite{Varlamov}.
Moreover the leading DOS contributions to the charge susceptibility 
cancel the MT ones. The AL diagrams give therefore 
the most important contribution to the charge susceptibility
(being the CT diagrams subdominant respect to them) \cite{notados}.

According to the physical assumption outlined above that 
the TDGL and KT temperature dependencies are essentially ruled by
the correlation lengths, we have alternatively taken Eq.(\ref{prop})
with $\xi=\xi_{GL}$ and $\xi=\xi_{KT}$.
In the calculation with $\xi_{GL}$ we used 
$T_c=T_{KT}$ and  the mass term $\epsilon=\ln(T/T_c)$, while in the
calculation with $\xi_{KT}$ we used Eq.(\ref{correlation}) 
with $b=1.6$ and $\xi_c=\xi_0$.
In both cases we took 
the coefficients $\eta$ and $\gamma$ given by the 
corresponding expressions reported below Eq.(\ref{prop})
calculated with $T_c=T_{BCS}$. 
This choice was motivated by the plausible assumption 
that $\eta$ and $\gamma$  change little once 
 the fluctuations are predominantly in the phase sector. 
In any case we checked that our results are rather stable with 
respect to  
modifications of $\eta$ and $\gamma$. 

The charge and the spin susceptibilities are finally  
obtained by the RPA resummation of the corrected charge and spin
bubbles
$
\chi _{c,s}={\chi _{c,s} ^{ph}}/{\left( 1 \pm (\tilde U_{c,s}/2)
\chi _{c,s}^{ph}\right)}
$
where the plus (minus) sign is associated to the charge (spin) 
susceptibility.
Notice that, following the analysis of Ref. \cite{Randeria}, 
the RPA expressions of both susceptibilities 
contain an effective local interaction $\tilde{U}_{c,s}$ instead of
the bare $U$ in order to properly fit the high temperature region
of the  Montecarlo data. 
The validity of the RPA form for the spin susceptibility is also found
in the context of the positive-$U$ Hubbard model \cite{Bulut}. 
However, while in Ref.\cite{Randeria} 
the bare bubbles were resummed and a value $\tilde{U}=6.5$ 
was obtained for $U=-4t$ and $<n>=0.5$, in our case we resum the
bubbles already containing the $\Delta \chi_s$ corrections and
a different value $\tilde{U}_s=-4.6$ is needed to match the
RPA calculation with the high temperature Montecarlo data.
For the charge susceptibility the comparison with the RPA resummation
in terms of the $\chi_0^{ph}$ reported in Fig. 2 gives
$\tilde{U}_c=-1.6$.

We now comment on the expression in Eq.1 that we used for the KT
correlation lenght. We wrote this expression following Halperin and 
Nelson \cite{Halperin}.
They introduce into the KT correlation length 
$\xi_{KT} \simeq a \exp \left[b(\pi J/k_B T-1) \right]$ for
the classical XY model (with coupling J and lattice spacing $a$) a
temperature dependent $J(T)=n_s(T)/8m$ and take $a=\xi_c$.
Here the superfluid density $n_s(T)$ is taken to vanish linearly 
at a temperature 
$T_0(>T_{KT})$ to be determined selfconsistently by the request that
$T_0$ should include the effect of the fluctuations at scale lower 
than $\xi_c$. Our expression (1) is obtained by taking
$T_0\simeq T_{BCS}$ and $\xi_c\simeq\xi_0$ with the idea that
phase fluctuations are the most important effect all over the range
of temperatures $T_{KT} \lesssim T \lesssim T_{BCS}$ 
(at least in evaluating $\chi_s$ and $\chi_c$)\cite{notakt}.

The results of the simple procedure outlined above
are quite satisfactory for the spin susceptibility. This indicates
that the main temperature dependence of this quantity actually arises
from the specific KT temperature dependence of the correlation length,
which thus brings along the physics of the vortex-antivortex phase
fluctuations into a simple perturbative scheme. 
The same is not true for the 
compressibility, as for the specific heat, 
since these quantities also involve temperature derivatives of $\xi_{KT}$.

Our method, developed for the 2D attractive Hubbard model, 
can be useful to understand the role of the 
superconducting phase fluctuations in quasi-2D cuprate superconductors.
In this context the recent finding that KT signatures, which 
are absent in the static conductivity, are progressively 
more evident in the 
dynamical conductivity at shorter timescales \cite{corson}
encourages to extend our analisys to other
frequency-dependent quantities.  In particular it is of
obvious interest to explore the possibility to 
include in a simple perturbative scheme along the lines followed in 
the present work the effects of KT topological phase 
fluctuations on dynamical quantities like
the optical conductivity and single-particle spectra.

{\bf Acknowledgments.} We acknowledge S. Caprara, C. Di Castro, 
P. Pieri, G. C. Strinati and A. A. Varlamov for helpful discussions.

\end{multicols}

\begin{thebibliography}{99}

\bibitem  {timusk} For a recent review see, {\it e.g.} T. Timusk and
	  B. Statt, Rep. Prog. Phys. {\bf 62}, 61 (1999).
\bibitem  {corson}J. Corson, R. Mallozzi, J. Orenstein, J. N. Eckstein,
          and I. Bozovic, Nature {\bf 398}, 221 (1999).
\bibitem  {Halperin} B. I. Halperin and D. R. Nelson, J. of Low T. Phys.
	  {\bf 36}, 599 (1979).
\bibitem  {KT1} J. M. Kosterlitz and D. J. Thouless, J. Phys. C {\bf 6}, 
	  1181 (1973).
\bibitem  {KT2} J. M. Kosterlitz, J. Phys. C {\bf 7}, 
	  1046 (1974). For a recent review see, Z. Gul\'acsi and M. 
	  Gul\'acsi, Advances in Physics {\bf 47}, 1 (1998).
\bibitem  {Randeria} M. Randeria {\em et al.}, Phys. Rev. Lett.
	  {\bf 69}, 2001 (1992).
\bibitem  {Singer} J. M. Singer {\em et al.}, 
	  Phys. Rev. B {\bf 54}, 1286 (1996).
\bibitem  {Trivedi} N. Trivedi and M. Randeria, Phys. Rev. Lett. {\bf 75},
          312 (1995).
\bibitem  {nota} We have consider a sligthly modified propagator which
	  has the periodicity of the lattice, substituting the $q^2$
	  term with $2(2-\cos q_x-\cos q_y)$; for small $q$ the two 
	  expressions are equivalent. 
\bibitem  {Micnas} For a review see, {\it e.g.} 
	  R. Micnas, J. Ranninger, and S. Robaszkiewicz,    
          Rev. Mod. Phys. {\bf 62}, 113 (1990).
\bibitem  {Varlamov} A. A. Varlamov, G. Balestrino, E. Milani, 
	  D. V. Livanov, Advances in Physics,
	  {\bf 48}, 6, 1 (1999).
\bibitem  {notados} Within an expansion of the bare density of states
	  with respect to the energy, the AL contribution is proportional
	  to $(1/N(\mu) \cdot \left . dN(E)/dE \right|_{E=\mu})^2$. 
	  This is in agreement with the
	  above discussion relating the AL contribution and $\chi_c$
	  to the dependence of the critical temperature on $\mu$.
\bibitem  {Bulut} N. Bulut {\em et al.}, Physica C {\bf 246}, 85 (1995).
\bibitem  {notakt} Our expression is at variance 
	  with respect to the one obtained in Ref. \cite{Nazarenko}
	  within a T-matrix selfconsistent approach. Their $\xi_{KT}$, 
	  rather strangely, would be obtained, in the 
	  context of Halperin and Nelson analysis, by assuming 
	  a linear increasing temperature dependence of the superfluid 
	  density for $T>T_{KT}$. 
\bibitem  {Nazarenko} J. R. Engelbrecht and A. Nazarenko, 
	  cond-mat/9806231, preprint (1998).  
\end{thebibliography}
\end{document}